\documentclass[aps,pre,twocolumn,superscriptaddress,showpacs]{revtex4-1}
\usepackage{graphicx}
\usepackage{array}
\usepackage{amsfonts}
\usepackage{amssymb,amsmath,multirow,rotate}
\usepackage{mathrsfs}
\usepackage{booktabs}
\usepackage{threeparttable}
\usepackage{multirow}
\usepackage{epsfig}
\usepackage{threeparttable}
\usepackage{chngpage}
\usepackage{latexsym}
\usepackage{dcolumn}
\usepackage{graphics,graphicx,bm,fleqn,epic,eepic,float}
\usepackage{verbatim}
\usepackage{subfigure}
\usepackage{color}

\definecolor{red}{rgb}{1,0,0}

\definecolor{blue}{rgb}{0,0,1}

\begin{document}

\title{Self adaptation of chimera states}
\date{\today}

\author{Nan Yao}
\affiliation{Department of Applied Physics, Xi'an University of Technology, Xi'an 710048, China}

\author{Zi-Gang Huang} \email{huangzg@xjtu.edu.cn}
\affiliation{School of Life Science and Technology, Xi'an Jiao Tong University, Xi'an 710049 China.}

\author{Hai-Peng Ren}
\affiliation{Shaanxi Key Laboratory of Complex System Control and Intelligent Information Processing, Xi'an University of Technology, Xi'an 710048, China}

\author{Celso Grebogi}
\affiliation{Institute for Complex Systems and Mathematical Biology, King's College, University of Aberdeen, Aberdeen AB24 3UE, United Kingdom}

\author{Ying-Cheng Lai}
\affiliation{School of Electrical, Computer and Energy Engineering, Arizona State University, Tempe, AZ 85287, USA}
\affiliation{Department of Physics, Arizona State University, Tempe, Arizona 85287, USA}

\begin{abstract}

Chimera states in spatiotemporal dynamical systems have been investigated
in physical, chemical, and biological systems, and have been shown to be
robust against random perturbations. How do chimera states achieve their
robustness? We uncover a self-adaptation behavior by which, upon a spatially
localized perturbation, the coherent component of the chimera state
spontaneously drifts to an optimal location as far away from the perturbation
as possible, exposing only its incoherent component to the perturbation to
minimize the disturbance. A systematic numerical analysis of the evolution
of the spatiotemporal pattern of the chimera state towards the optimal stable
state reveals an exponential relaxation process independent of the spatial
location of the perturbation, implying that its effects can be modeled
as restoring and damping forces in a mechanical system and enabling the
articulation of a phenomenological model. Not only is the model able to
reproduce the numerical results, it can also predict the trajectory of
drifting. Our finding is striking as it reveals that, inherently, chimera
states possess a kind of ``intelligence'' in achieving robustness through
self adaptation. The behavior can be exploited for controlled generation of
chimera states with their coherent component placed in any desired spatial
region of the system.

\end{abstract}

\maketitle

In spatially extended nonlinear dynamical systems, spontaneous symmetry
breaking is common. For example, in a system of nonlocally coupled,
identical nonlinear oscillators, coexistence of coherence and incoherence in
distinct spatial regions can emerge during the spatiotemporal evolution of
the system. This remarkable phenomenon was first observed about three decades
ago in a numerical study of a system of coupled nonlinear Duffing
oscillators~\cite{UGOA:1989}, and was termed as ``domain-like spatial
structure.'' Later, the phenomenon was rediscovered~\cite{KB:2002},
analyzed and given the name of ``chimera''~\cite{AS:2004,SK:2004}. Since then,
there has been a great deal of interest in the
subject~\cite{AS:2006,AMSW:2008,SSA:2008,Laing:2009a,Laing:2009b,SCL:2009,
Martens:2010,MLS:2010, OWM:2010,WO:2011,WOYM:2011,OMHS:2011,OWYMS:2012,
ZLZY:2012,LRK:2012,TNS:2012,HMRHOS:2012,OOHS:2013,UR:2013,ZZY:2013,YHLZ:2013,
NTS:2013,MTFH:2013,LBM:2013,PA:2013,GSD:2013,SOW:2014,SSKG:2014,ZZY:2014,
O:2013,OMT:2008,XKK:2014,YHGL:2015,MSOM:2015,PA:2015,PAb:2015,MB:2015,
OZHSS:2015,NTS:2016,VA:2016,HBMR:2016,GF:2016,SZMS:2016,KHSKK:2016,UOZS:2016,
ARM:2017,BMGP:2017,RBPG:2017,SSAZ:2017,MOSH:2018,BK:2018,OOZS:2018,XWHL:2018}.
Chimera states have been studied in different types of systems such as
regular networks of phase-coupled oscillators with a ring
topology~\cite{KB:2002,AS:2004,AS:2006}, regular networks hosting a few
populations~\cite{AMSW:2008,Martens:2010},
two-dimensional~\cite{SK:2004,MLS:2010} and three-dimensional
lattices~\cite{MSOM:2015}, torus~\cite{PA:2013,OWYMS:2012}, and systems with a
spherical topology~\cite{PA:2015}. Issues that were addressed include
transient behaviors associated with chimera
states~\cite{OWM:2010,WO:2011,WOYM:2011}, the
effects of time delay~\cite{SCL:2009,SSA:2008,OMT:2008}, phase
lags~\cite{ZLZY:2012}, coupling functions~\cite{OOHS:2013,UR:2013,ZZY:2013},
and the impacts of random perturbation and complex topology of
coupling~\cite{LRK:2012,YHLZ:2013,ZZY:2014,YHGL:2015}.
Experimentally, chimera states have been observed in a system of chemical
oscillators~\cite{TNS:2012,NTS:2013,NTS:2016},
in an optical system~\cite{HMRHOS:2012,HBMR:2016},
in coupled mechanical oscillators~\cite{MTFH:2013}, in electrochemical
systems~\cite{LBM:2013,SSKG:2014}, and even in quantum
systems~\cite{VA:2016,XWHL:2018}. Natural phenomena associated with chimera
states include unihemispheric sleep~\cite{RAL:2000,MWL:2010}, neural
spikes~\cite{Sakaguchi:2006,OPT:2010}, and possibly ventricular
fibrillations~\cite{DPSBJ:1992}. Control of chimera states has also been
investigated~\cite{SOW:2014,MB:2015,SZMS:2016,GF:2016,OOZS:2018}.

An issue of both theoretical and experimental interest is the robustness
of the chimera states against external perturbations. In this regard,
the effects of random removal of links were studied~\cite{YHLZ:2013} with the
finding that, even when a large number of links are removed so that chimera
states are deemed not possible, in the state space there are still both
coherent and incoherent regions, and the regime of conventional chimera state
is a particular case in which the oscillators in the coherent region happen to
be synchronized or phase-locked. Another work on networks of FitzHugh-Nagumo
oscillators demonstrated that the chimera states are robust against irregular
structural perturbations~\cite{OPHSH:2015}. Quite recently, the robustness of
chimera states in nonlocally coupled networks of nonidentical logistic maps
was investigated~\cite{MOSH:2018}. These studies indicate that chimera states
are generally robust against various kinds of external perturbations. The
question is how does a chimera state respond to perturbation to achieve its
robustness. Specifically, suppose the coherent component of the chimera
state is disturbed so that the component is no longer coherent. If the state
is to survive, it must adjust the relative distribution of the coherent and
incoherent components in the space. That is, upon perturbation, a chimera
state must reorganize itself into a new state, possibly through self
adaptation, to generate a modified distribution of the coherent and incoherent
components. How does the system accomplish this feat?

In this paper, we report a remarkable phenomenon of self adaptation of
chimera states. When a spatially localized external perturbation is applied
to the coherent component of a chimera state, it initiates and executes a
self-adaptive drifting process toward an optimal state in which the incoherent
component masks the perturbation and the newly formed coherent component is
as far away as possible from the perturbation site. The response of the
system is then to evolve toward a new chimera state that shields itself from
the perturbation in an optimal way. Not only that, the system is also capable
of selecting the optimal path towards the new chimera state. Carrying out a
detailed analysis of the collective dynamics and patterns associated with the
spatiotemporal evolution of the chimera, we identify the essential physical
ingredients associated with the self-adaption process: an exponential
relaxation of the chimera state toward the new stable state and the collapse
of the relaxation trajectories into a single one independent of the location
of perturbation. These behaviors enable us to construct a phenomenological
model for a physical understanding of self adaptation of the chimera states.
Taken together, the response of a chimera state to a perturbation through
self adaptation is indicative of some intrinsic ``intelligence'' of the state,
which not only is theoretically interesting, but also has implications to
control or manipulation of chimera states in experimental systems.

\begin{figure*}[htb!]
\includegraphics[width=0.9\linewidth]{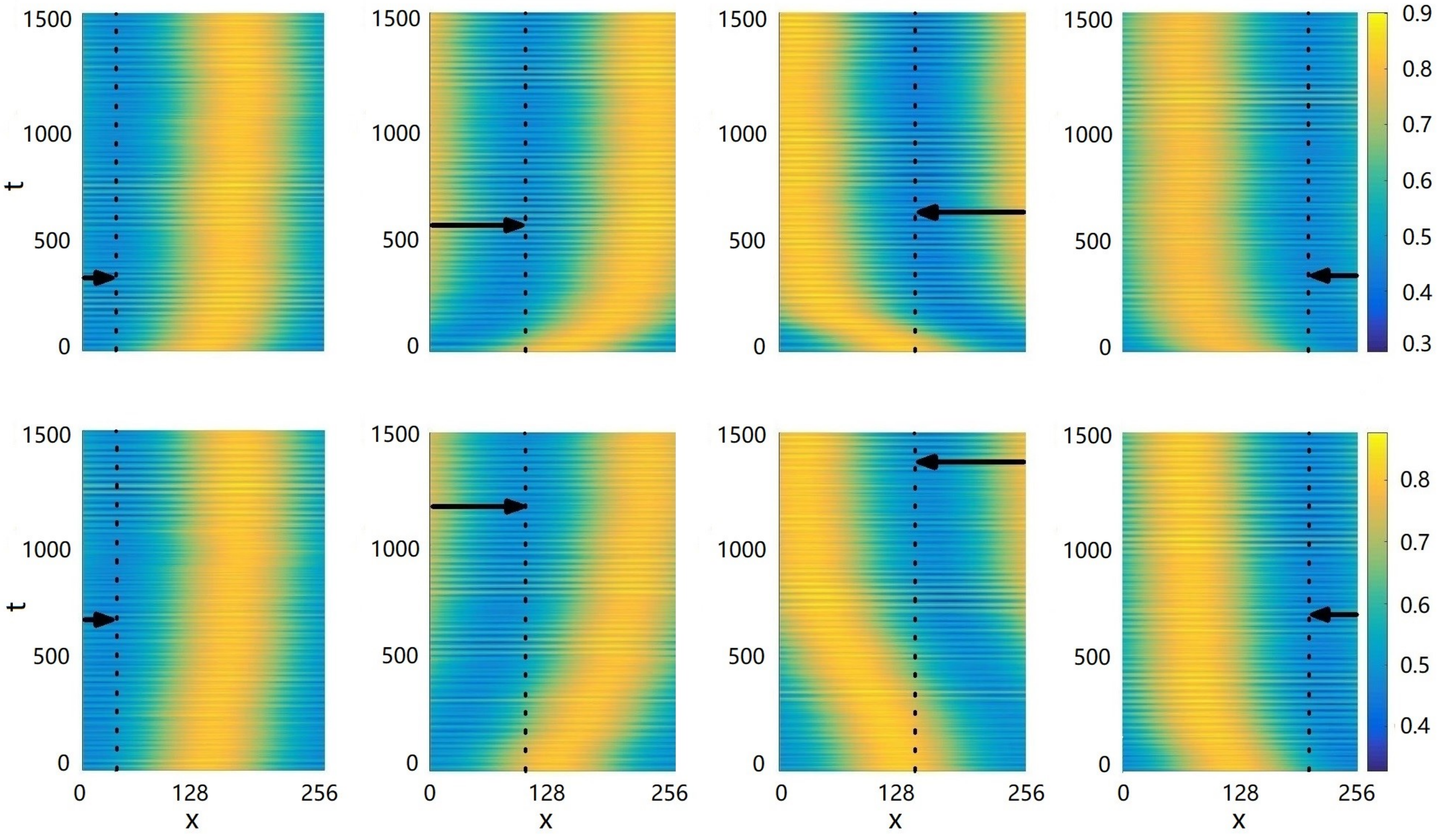}
\caption{ (Color online) Self-adaptive, ``intelligent'' drift of chimera
state in response to a spatially localized perturbation. The coherent and
incoherent regions are represented by the yellow and blue colors,
respectively. The drift is activated by disturbing a single node at the
location $x_0$, where the perturbation strength is $\Delta\phi=0.3\pi$ for
the upper panels and $0.9\pi$ for the lower panels. In each panel, the black
arrow indicates the direction of drifting of the chimera state and the arrow
length represents the drifted distance. The vertical location of the arrow
specifies the time when the drifting chimera state settles down (or becomes
stable). Other parameters are $A = 0.995$, $\alpha = 1.39$, and $N=256$
(system size).}
\label{fig:pattern}
\end{figure*}

We consider the paradigmatic setting for studying chimera
states~\cite{KB:2002,AS:2004,AS:2006}: a ring network of $N$ non-locally
coupled, identical phase oscillators with the periodic boundary condition:
$d\phi(x_i)/dt = \omega - (2\pi/N)\sum_{j=1}^N c_{ij}G(x_i,x_j)
\sin{[\phi(x_i) - \phi(x_j) + \alpha]}$, where $\phi(x_i)$ is the phase of
the $i$th oscillator at spatial location $x_i$ and the range of the spatial
variable is $[-\pi,\pi]$. The angular velocity $\omega$ and phase lag $\alpha$
of the oscillators are constants in space. Without loss of generality, we
set $\omega = 0$ and $\alpha \alt \pi/2$. The kernel
$G(x_{i},x_{j}) = [1+A\cos{(x_{i}-x_{j})}]/(2\pi)$
is a non-negative even function that defines the non-local coupling
among all the oscillators. The quantity $c_{ij}$ is the $ij$th element of
the $N\times N$ coupling matrix ${\bf C}$, where $c_{ij} = 1$ if there is
coupling from the $j$th oscillator to the $i$th oscillator, and $c_{ij} = 0$
indicates the absence of such coupling. For the ring system, chimera states
are common~\cite{KB:2002,AS:2004,AS:2006}, as exemplified in
Fig.~\ref{fig:pattern}.

To assess how perturbations affect the chimera state, we disturb the dynamical
variable of a single oscillator (the target oscillator) at location $x_0$ that
belongs to the coherent component. The nature of the perturbation is to force
upon the oscillator a constant phase difference $\Delta \phi$ with respect to
the local mean phase $\overline{\phi}_{\mathrm{local}}$ of its $2z$ neighbors,
with equal number of neighbors on the left and right sides. Because of the
perturbation, the originally coherent component is no longer coherent, and the
chimera state, if it is to remain, must adjust itself to a new stable state.
How does this occur?

Figure~\ref{fig:pattern} shows the final spatiotemporal pattern of the chimera
state in response to perturbation of two strength values at different
locations (indicated by the vertical dashed lines in different panels).
Instead of evolving into a globally coherent or incohereint state, the
original state maintains its chimerical character by shifting the coherent
component to a new region that is as far away as possible from the perturbed
oscillator. At the same time, the incoherent component evolves to a region
that contains the perturbed oscillator approximately at the center. This
remarkable self adaptive behavior represents an ``intelligent'' scheme of
the chimera state to protect itself. Numerically, we also observe that,
a perturbation to an oscillator in the originally incoherent region tends
to expand the incoherent region but only slightly, so the effects on the
chimera state are much less dramatic than those with perturbation in the
coherent region.

Two characteristics of the spatiotemporal evolution of the chimera state
in response to a perturbation are as follows. Firstly, after the perturbation
is applied at $x_0$, the incoherent region begins to drift until its center
$x_\mathrm{mid}(t)$ reaches $x_0$. This is surprising as, intuitively, one
might expect the drift to stop once the incoherent region contains the
location $x_0$. Each panel in Fig.~\ref{fig:pattern} presents the relevant
features: the midpoint $x_\mathrm{mid}(t)$ of the incoherent (blue) region,
the target node at $x_0$, the corresponding time for $x_\mathrm{mid}(t)$ to
reach $x_0$ (the vertical location $\tau$ of the arrow), and the instant when
$x_0$ is just covered by the incoherent region (indicated by
$t_{\mathrm{cover}}$ at which the coherent-incoherent boundary
$x_{\mathrm{bound}}$ crosses $x_0$). We have $t_{\mathrm{cover}}<\tau$,
indicating that the drift is not terminated even when the target node has
already been covered by the incoherent region. The phenomenon is
counterintuitive because the expectation is that, once the target node is
merged in the incoherent region, the movement of the state should stop as
the phases and the velocities of the individual oscillators in thei
incoherent region are nonetheless intrinsically random. The fact that the
state continues to drift until $x_\mathrm{mid}$ reaches $x_0$ implies
a kind of self adaptation among the oscillators toward an optimal state
that makes the chimera state as robust as possible. Indeed, the drift
terminates when $x_\mathrm{mid} = x_0$ so that the new chimera state
possesses a global symmetry with maximum robustness. Because of the ``desire''
for the chimera state to acquire the symmetry, a perturbation even in
the originally incoherent region, which breaks the global symmetry of the
chimera state, would induce a drift. This has indeed been observed
numerically. In fact, once the state has been stabilized, the order
parameter $R(x)$ of the midpoint $x_\mathrm{mid}$ in the incoherent region
reaches a minimum value, providing a way to calculate the value of
$x_\mathrm{mid}$. Secondly, the system always chooses the shorter path for
$x_\mathrm{mid}$ to drift toward $x_0$, as indicated by the length of
the arrow in each panel of Fig.~\ref{fig:pattern}. Especially, because of
the periodic boundary condition, there are two possible routes of drift.
For every case examined, the drift takes place along the shorter path.

\begin{figure}[htb!]
\includegraphics[width=\linewidth]{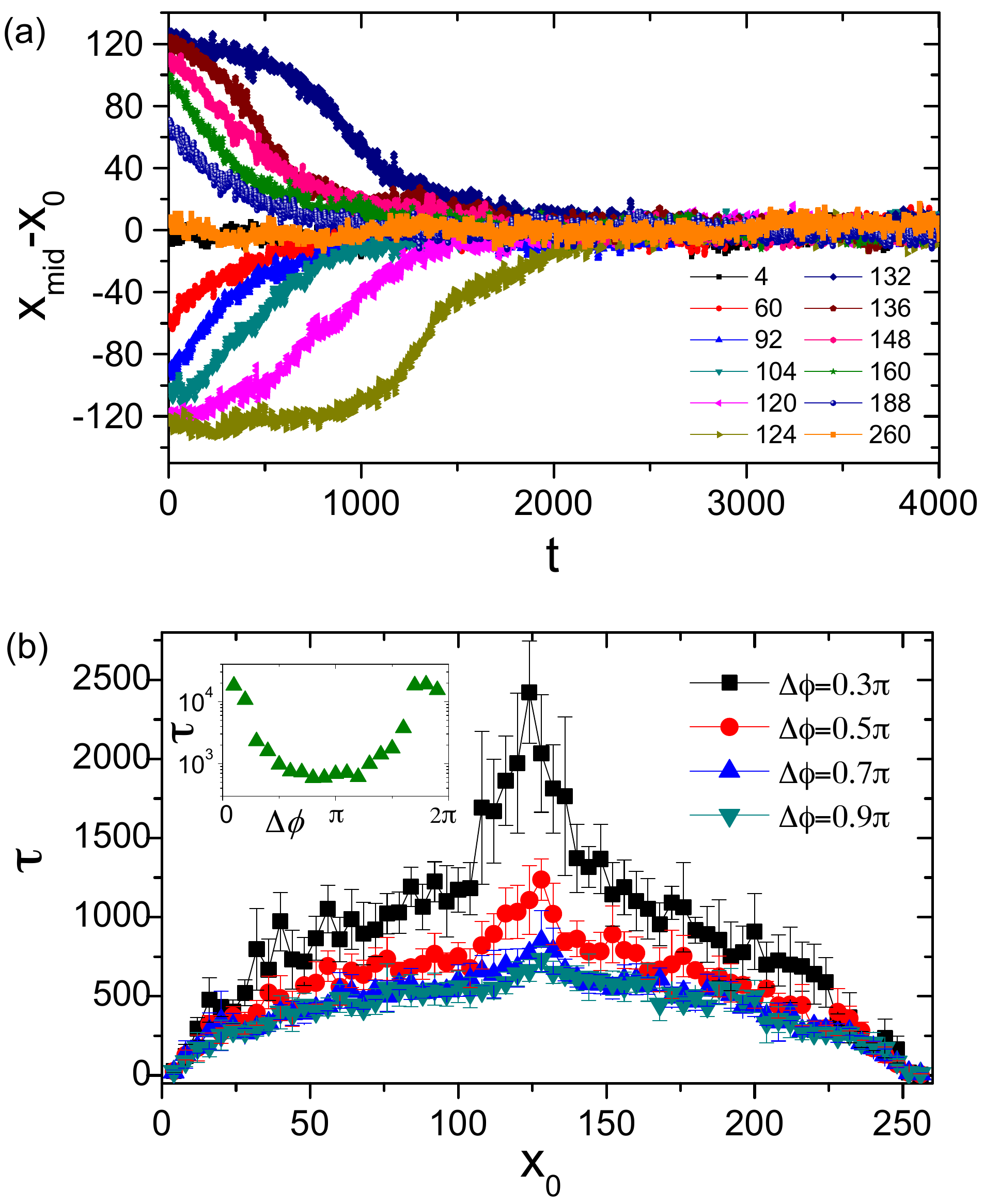}
\caption{ (Color online) Spatial movement and relaxation time of chimera
associated with self-adaptive drifting. (a) Distance between the middle
point $x_{\mathrm{mid}}$ of the incoherent region of the chimera state and
the location $x_0$ of the target oscillator. For different values of $x_0$,
the chimera state drifts until $x_\mathrm{mid}(t)$ covers $x_0$, i.e.,
$x_\mathrm{mid}(t)-x_0$ converges to $0$. (b) The relaxation time $\tau$
for the chimera state to become stable again versus $x_0$ and the
perturbation strength $\Delta\phi$, respectively.}
\label{fig:DeltaxTau}
\end{figure}

To gain further insights into the physical mechanism of the self-adaptive
behavior of chimera states, we examine the temporal dynamics of drifting.
Specifically, for a given chimera state with its coherent region centered at
$N/2$, we monitor the evolution of
$\Delta x(t) \equiv |x_\mathrm{mid}(t) - x_0|$ for different values of $x_0$,
as shown in Fig.~\ref{fig:DeltaxTau}(a) for $\Delta\phi=0.3\pi$. In all cases,
$\Delta x(t)$ converges to zero with small fluctuations. The relaxation time
$\tau$ of the self-adaptive drifting is effectively the first passage time
of the smoothed $\Delta x(t)$ curve to zero. Figure~\ref{fig:DeltaxTau}(b)
shows $\tau$ versus $x_0$ and $\Delta\phi$. We see that, when $x_0$ is closer
to the center of the coherent region ($x=N/2$), $x_\mathrm{mid}(t)$ travels a
longer distance to reach $x_0$, leading to a larger value of $\tau$.
The impact of the perturbation strength $\Delta\phi$ on the drifting process
is symmetric about $\pi$ under the periodic boundary condition, as shown in
the inset of Fig.~\ref{fig:DeltaxTau}(b). It can also be seen that, for
small perturbation ($\Delta\phi \sim 0$ or $2\pi$), the drifting process
slows down significantly with the relaxation time $\tau$ approximately one
order of magnitude higher than that associated with $\Delta\phi \sim \pi$.

\begin{figure}[htb!]
\includegraphics[width=\linewidth]{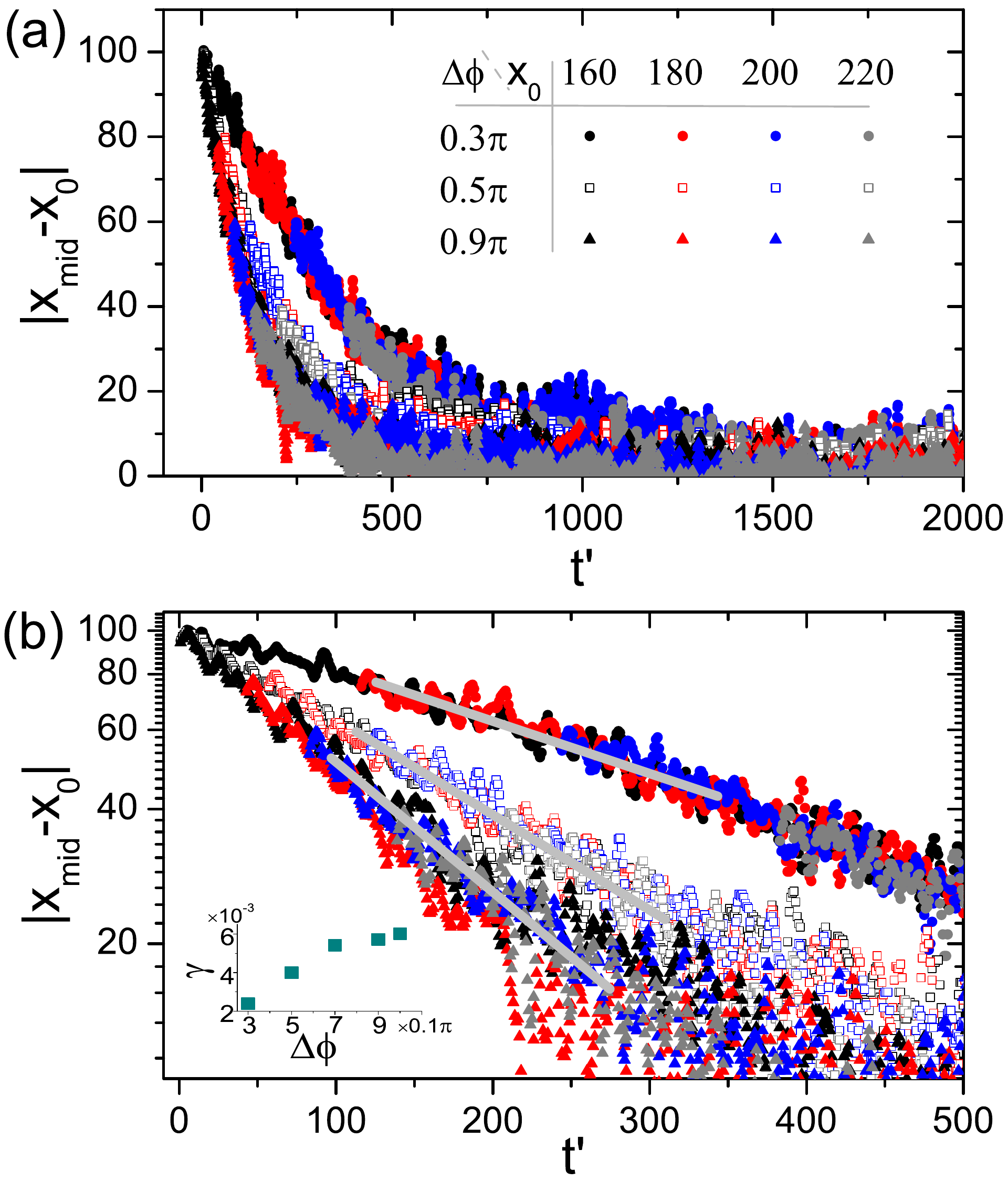}
\caption{ (Color online) Collapse of drifting trajectories. (a) In the
transformed coordinate $t'$, the absolute distances $|x_\mathrm{mid}(t')-x_0|$
resulting from different values of $x_0$ collapse into one, implying that
the chimera state drifts according to the distance between $x_\mathrm{mid}(t)$
and $x_0$. (b) The same data on a logarithmic-normal plot, where the
exponentially decaying behavior $|x_\mathrm{mid}(t')-x_0|\sim e^{-\gamma t}$
can be seen. The perturbation strength is $\Delta\phi=0.3\pi$ (solid circles),
$0.5\pi$ (open squares), and $0.9\pi$ (solid triangles). The value of the
exponential rate $\gamma$ increases with $\Delta\phi$.}
\label{fig:DeltaxAbs}
\end{figure}

Does the self-adaptive drifting process have any memory of the value of $x_0$?
The question can be addressed by examining whether two intermediate states
evolving from different initial states and having the same value of
$\Delta x(t)$ at some time $t$ can be distinguished. To facilitate a
comparison, we use the transformed time $t'=t+t_{0}(x_0)$, where $t_{0}(x_0)$
is the time at which the dynamical variables of all the oscillators collapse to
a single point. Any subsequent collapse would be indicative of lack of any
memory effect. Figure~\ref{fig:DeltaxAbs}(a) shows $|\Delta x(t')|$ for
different values of $x_0$. The three classes of collapsed curves correspond
to different values of the perturbation strength $\Delta\phi$. Because of
the collapses, any memory effect in the spatiotemporal evolution of the
chimera state upon perturbation can be ruled out.
Figure~\ref{fig:DeltaxAbs}(b) shows the same data but on a logarithmic-normal
plot, which indicates an exponential decay: $\Delta x(t)\sim e^{-\gamma t}$,
with $\gamma$ being the rate of decay whose value increases with $\Delta\phi$.
That is, a larger perturbation induces faster drifting of the chimera.
Note that, for a given value of the perturbation strength, all the
trajectories collapse into one, indicating that the distance between
$x_\mathrm{mid}(t)$ and $x_0$ is the sole factor determining the
self-adaptive drifting process.

To gain theoretical insights, we examine the effect of a particular type of
perturbations: these applied to oscillators at the boundaries between the
coherent and incoherent regions located at $x_{\mathrm{bound}}$, as the
drifting process is essentially determined by the movement of the boundaries.
Let $\eta=n_{\mathrm{incoh}}/N$ be the fraction of the incoherent region
associated with an unperturbed chimera state, where the value of
$\eta$ depends on parameters such as the coupling strength $A$ and the
phase lag $\alpha$. When a perturbation is applied to the oscillator at the
center of the incoherent region, the value of $\eta$ tends to increase
slightly, somewhat pushing the boundary into the coherent region. However,
analysis reveals that any small change in the value of $\eta$ tends to
diminish, restoring the original ratio between the coherent and incoherent
regions~\cite{AS:2006}.

\begin{figure}[htb!]
\includegraphics[width=\linewidth]{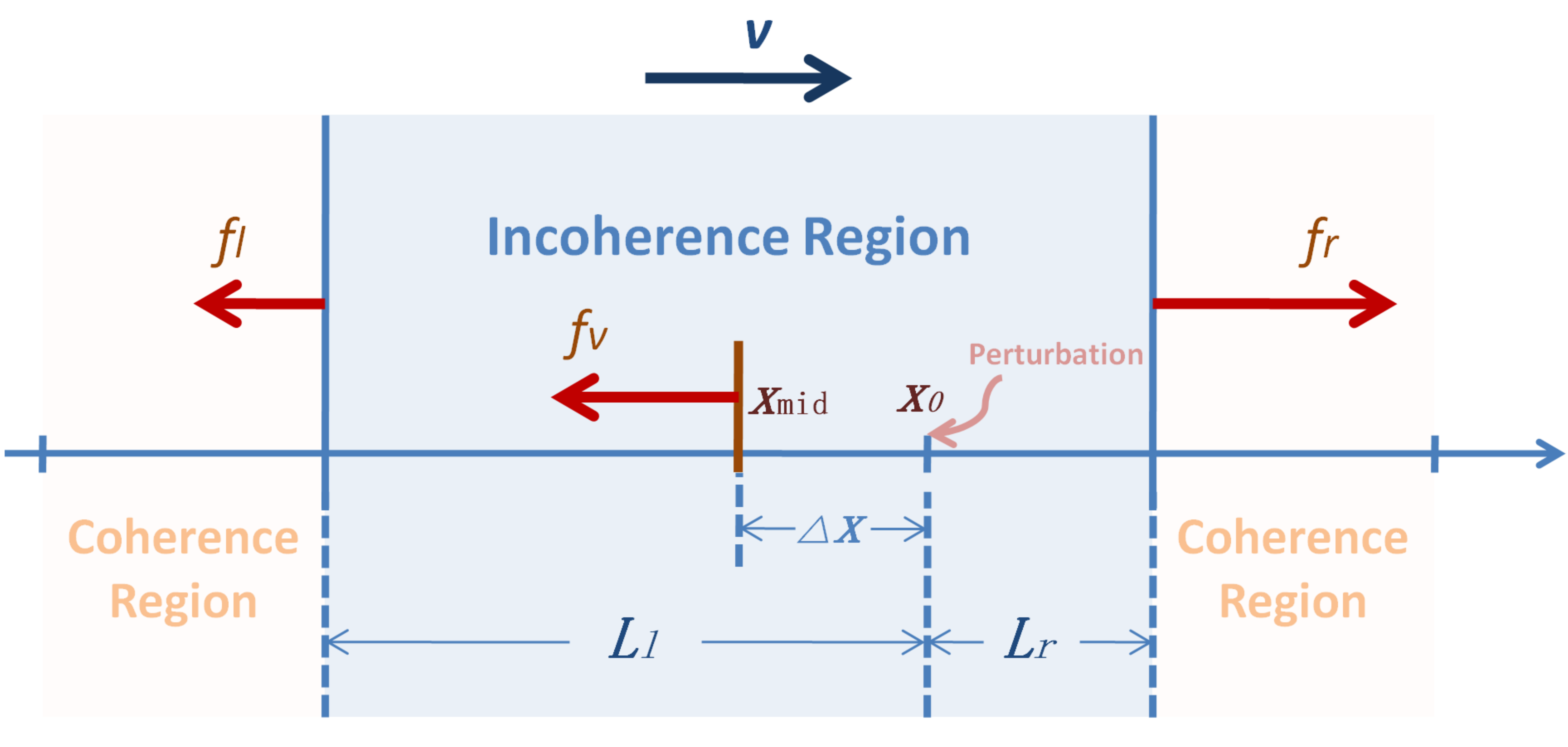}
\caption{(Color online) A schematic illustration of the effective,
perturbation-induced forces in the phenomenological model. Perturbation at
$x_0$ induces forces $f_l$ and $f_r$ on the left and right coherent-incoherent
boundaries, respectively. The chimera state drifts towards the right-hand side,
indicating the existence of a damping force $f_v$}
\label{fig:sketchmap}
\end{figure}

Based on the numerical results, we articulate a phenomenological model to
account for the impact of perturbation on the chimera state.
Figure~\ref{fig:sketchmap} presents a schematic illustration of the dynamics
of the boundaries between the coherent and incoherent regions, where the left
and right boundaries are located at at $x_l$ and $x_r$, respectively. Let
$f_l$ and $f_r$ be the effective forces induced by the perturbation at $x_0$
to push the left and right boundaries, respectively. The distance from $x_0$
to the left (right) boundary is $L_l$ ($L_r$), and the width of the
incoherent region is $L= L_l + L_r $. The effective force $f_l$ ($f_r$)
depends on $L_l$ ($L_r$). The mathematical forms of these forces can be
derived from the dynamical behavior of $\Delta x(t)$. In particular, the
exponential decay of $\Delta x(t)$ with time indicates that the velocity
and acceleration of the drifting also decay exponentially with time at the same
rate. We define $y\equiv |\Delta x(t)| \mathcal{A}e^{-\gamma t}$ to obtain
$\dot{y}=-\gamma \mathcal{A} e^{-\gamma t}=-\gamma y$ and
$\ddot{y}=\gamma^2 \mathcal{A} e^{-\gamma t}=\gamma^{2}y$. The effective
force upon the chimera state can be written as $F=m\ddot{y}=m\gamma^{2}y$.

The linear dependence of the effective force $F$ on $y$ suggests that the
force contain two components: a linear restoring force $F_k(y) = - k y$ and
a damping force $f_v = -\eta \dot{y}$, with $k$ being the elastic
constant and $\eta$ being the damping coefficient. The evolution of
$y(t)$ obeys the equation: $m\ddot{y}+\eta \dot{y} + k y = 0$.
From the function of $\ddot{y}$ and $\dot{y}$, we have
$m\gamma^{2}y - \eta\gamma y + ky =0$, leading to the relation
$k = \gamma (\eta - m\gamma)$ and hence the critical value of damping
beyond which $y(t)$ decays exponentially to zero. The effect of perturbation
on the chimera state can then be regarded as the result of the forces acting
upon the two boundaries: $F_k(y)=f_r+f_l$. We have
$F_k(y)=  -\gamma (\eta - m\gamma) y$. Since $y=(L_l - L_r)/2$, we can also
get the forces acting upon the left and right boundaries as
$f_l = -\mathcal{B} \gamma^{2} (L_0-L_l)/2$ and
$f_r = \mathcal{B} \gamma^{2} (L_0-L_r)/2$, respectively. The value of $L_0$
does not affect the movement of the chimera state. Because of the
conservative nature of the restoring force, the minimum potential energy
occurs at $y=0$. The presence of the critical linear damping force $f_v$
leads to the exponential decay of $y(t)$ towards the minimum energy state.
The phenomenological model thus explains the perturbation induced,
self-adaptive drifting dynamics of the chimera state.

To further justify the phenomenological model, we resort to the two
commonly used theoretical tools in the analysis of chimera states: the
continuity equation~\cite{Laing:2009b} and the concept of invariant
manifold~\cite{OA:2008,OA:2009}. In general, the chimera dynamics can be
characterized~\cite{KB:2002} by the following complex order parameter
$Z$ defined for oscillator $i$ as
$Z(x_{i})\equiv R({x_i}){e^{\mathrm{i}\Theta ({x_i})}} =
(2\pi/N)\sum_{j=1}^N {G({x_i} - {x_j})e^{\mathrm{i}\theta ({x_j})}}$,
where the phase of the oscillator is $\theta=\phi-\Omega t$ with $\Omega$
being the phase velocity of the oscillators in the coherent subset when
a chimera state emerges. Theoretical insights into the chimera states
can be obtained by examining the continuum limit $N\rightarrow\infty$,
where the system can be described by a one-dimensional
PDE~\cite{OA:2008,OA:2009}. In particular, the state of the system can be
characterized by a probability density function $f(x,\phi,t)$ governed
by the continuity equation
$\partial f/\partial t + \partial/\partial \phi (f\mathrm{v} ) = 0$,
with $\mathrm{v}$ being the phase velocity~\cite{Laing:2009b}. The function
$f(x,\phi,t)$ can be expressed in terms of Fourier series expansion as
$f(x,\phi,t)= [1/(2\pi)]\{1+\sum_{n = 1}^\infty
{[h^n(x,t) e^{\mathrm{i}n\phi} + \mbox{c.c.}]}\}$,
where ``$\mbox{c.c.}$'' stands for the complex conjugate of the preceding
term, and the $n$th coefficient is the $n$th power of some function $h(x,t)$
that effectively characterizes the state of the system. The time evolution
of $h(x,t)$ associated with the order parameter $Z(x,t)$ is given
by~\cite{OA:2008,OA:2009}
$\partial h(x,t)/\partial t = -{\mathrm{i}}\omega h(x,t) +
\frac{1}{2}\left[ {{Z^*(x,t)}{e^{{\mathrm{i}}\alpha }}
- Z(x,t){e^{ -{\mathrm{i}}\alpha }}{h^2}(x,t)} \right]$, where
$Z(x,t) =  \int_{-\pi}^{\pi}{G(x-x')h^*(x',t)dx'}$ and
$G(x-x')$ is the coupling function with normalized $x$:
$G(x-x')=[1+A\cos{(x-x')}]/(2\pi)$ for $-\pi < |x-x'| \leq \pi$.
Since the perturbation $\Delta \phi$ upon the phase of one single oscillator
does not break the spacial pattern of the chimera but just induces the
drifting of chimera as a whole, the theoretical description is applicable.

The impact of perturbation $\Delta \phi$ at $x_0$ can be characterized as
$h(x_0)=h_0(x_0)e^{i\Delta \phi}$ or
$h^{\ast}(x_0)=h^{\ast}_0(x_0)e^{-i\Delta \phi}$ based on the Fourier
series expansion, with $h_0(x_0)$ and $h^{\ast}_0(x_0)$ denoting the
respective values in the absence of perturbation. We then have
$\delta h(x_0)=h_0(x_0)(e^{i\Delta \phi}-1)$ and
$\delta h^{\ast}(x_0)=h^{\ast}_0(x_0)(e^{-i\Delta \phi}-1)$.
From the evolutionary equation of $h(x,t)$, we have that the variances of
$Z$ and $Z^{\ast}$  due to the perturbation at $x_0$ are
$\delta Z=(2\pi/N)G(x-x_0)\delta h^{\ast}(x_0)$ and
$\delta Z^\ast=(2\pi/N)G(x-x_0)\delta h(x_0)$, respectively. The variance
of $\partial h/\partial t$ is
$\delta \dot{h}(x,t)=-(1/2)e^{-i\alpha}h^2(x,t)\delta Z +(1/2)e^{i\alpha}\delta Z^{\ast}=(\pi/N)G(x-x_0)[\mathcal{X}-\mathcal{X}^{\ast}h^{2}(x,t)]$,
with $\mathcal{X}=e^{i\alpha} \delta h(x_0)$. This provides a physical picture
of how perturbation affects the chimera state. In particular, a larger value
of the perturbation strength $\Delta \phi$ leads to a higher probability for
a larger deviation $\delta \dot{h}(x,t)$ in the evolution, and the focal
oscillator at $x$ with a smaller distance to $x_0$ gains a larger value of
$\delta \dot{h}$ due to the larger value of $G(x-x_0)$. The deviation
$\delta \dot{h}$ from the original chimera state reduces the stability of
the coherent region and enlarges the incoherent region from the boundaries
of the two regions at the speed $\delta \dot{h}(x,t)$. As shown in
Fig.~\ref{fig:sketchmap}, a larger disturbance takes place at the oscillator
closer to the boundary, i.e., the right-hand side boundary at $x_r$ (since
$L_r <L_l$). Additionally, due to the intrinsic inertia of the chimera state
to maintain the fraction between the coherent and incoherent regions, the
expansion of incoherent region takes place at the right-hand side boundary.

To summarize, we uncover a striking phenomenon that occurs when a chimera
state is disturbed: the state is capable of self-organizing into a new stable
state in an adaptive and optimal way. Especially, when a spatially localized
perturbation is applied to the coherent region, the chimera state is able to
quickly ``move'' in the space (in an exponential fashion) to generate a new
stable coherent region at the maximum distance from the perturbed oscillator
through a path that is energy efficient. All these happen as if the chimera
was ``intelligent.'' We develop a simple mechanical model to account for these
features, which is justified qualitatively by a theoretical analysis. It has
been known that chimera states are robust. Our work provides a clear physical
and dynamical picture on how the robustness is achieved. Experimental effort
to verify the self-adaptive dynamics of chimera states iuncovered in this
paper will be appreciated.


%

\end{document}